\documentstyle[seceq,short,twocolumn]{jpsj}

\newcommand{\mathbd}[1]{\mbox{\boldmath ${#1}$}}

\def\runtitle{}
\def\runauthor{Akihiro {\sc Ishizawa} and Yuji {\sc Hattori}}

\title{Large Coherent Structure Formation by Magnetic Stretching Term
in Two-Dimensional MHD Turbulence}

\author
{ 
Akihiro {\sc Ishizawa}\footnote{
E-mail: ishizawa@fusion.naka.jaeri.go.jp}
and Yuji {\sc Hattori}$^{1,}$\footnote{
Present address:
Faculty of Engineering, Kyushu Institute of Technology, 
Kitakyushu 804-8550.
E-mail: hattori@mns.kyutech.ac.jp
}
}

\inst
{
Naka Fusion Research Establishment, Japan Atomic Energy
Research Institute, Ibaraki 311-0193 \\
$^1$Institute of Fluid Science, Tohoku University, Sendai 980-8577 \\
}

\recdate {June 8, 1998}

\kword
{
MHD turbulence, self-organization, coherent structure,
EDQNM, magnetic stretching term, eddy viscosity
}

\begin{document}
\sloppy
\maketitle

In the 2-D MHD turbulence there arise large coherent magnetic 
structures.\cite{Biskamp1}
The formation of these coherent structures is 
interpreted as the self-organization due to the selective decay and 
the dynamic alignment\cite{Biskamp1};  
it is related to the inverse cascade and
understood as the coalescence of small magnetic eddies
due to the high eddy viscosity
(turbulent diffusivity) in recent papers.\cite{Biskamp2}\cite{Kinney}
In the 2-D non-MHD turbulence the coherent vortical structures are known 
to form.\cite{Brachet}
In this case, since there is only one nonlinear term, 
$\mathbd{v} \cdot \nabla \mathbd{v}$, 
this convection term should be responsible for the formation 
of coherent structures. 
In the 2-D MHD turbulence, however, 
since we have several nonlinear terms in the equations of motion, 
it is not clear which term contributes to the formation of 
coherent structures. 
Our previous numerical results\cite{Ishizawa} have suggested that 
the magnetic stretching term conveys the energy from small scale to 
large scale, which implies that this term is responsible for 
the formation of coherent structures. 

In order to elucidate
which term contributes to the formation of large coherent magnetic structures,
we investigate the roles of nonlinear terms in the MHD equations
by estimating the eddy viscosity 
with the eddy damped quasi-normal Markovian (EDQNM) 
approximation.\cite{Orszag}
The eddy viscosity represents the effect of the small-scale
turbulence on the large-scale velocity and magnetic fields 
and its sign implies the direction of the energy cascade.
It will be shown that the sign of the eddy viscosity
due to the magnetic stretching term,
$\mathbd{B} \cdot \nabla \mathbd{v}$, 
is negative and thus this term contributes to
the formation of the large coherent magnetic structures.

We consider the homogeneous isotropic turbulence
of incompressible conducting fluid.
The non-dimensionalized equations for an MHD flow are written as
\begin{equation}
{{\partial \mbox{\boldmath $v$}} \over {\partial t}}+\mbox{\boldmath $v$}
\cdot \nabla \mbox{\boldmath $v$}=-\nabla p_t+\mbox{\boldmath $B$}\cdot 
\nabla \mbox{\boldmath $B$}+\nu \Delta \mbox{\boldmath $v$},
\label{eq:NS}
\end{equation}
\begin{equation}
{{\partial \mbox{\boldmath $B$}} \over {\partial t}}+\mbox{\boldmath $v$}
\cdot \nabla \mbox{\boldmath $B$}=\mbox{\boldmath $B$}\cdot \nabla 
\mbox{\boldmath $v$}+\eta \Delta \mbox{\boldmath $B$},
\label{eq:Maxwell}
\end{equation}
\begin{equation}
\nabla \cdot \mbox{\boldmath $v$}=0,
\qquad \nabla \cdot \mbox{\boldmath $B$}=0,
\label{eq:div}
\end{equation}
where $\mbox{\boldmath $v$}$, \mbox{\boldmath $B$} and $p_t$ denote 
the velocity field, the magnetic field and the total pressure, respectively,  
and $\nu$ and $\eta$ denote the kinematic viscosity and the 
magnetic diffusivity.
We call the nonlinear terms 
$\mbox{\boldmath $v$}\cdot \nabla \mbox{\boldmath $v$}$, 
$\mbox{\boldmath $B$}\cdot \nabla \mbox{\boldmath $B$}$, 
$\mbox{\boldmath $v$}\cdot \nabla \mbox{\boldmath $B$}$  
and $\mbox{\boldmath $B$}\cdot \nabla \mbox{\boldmath $v$}$ 
as hydrodynamic convection, magnetic tension, magnetic convection 
and magnetic stretching, respectively. 
Note that the hydrodynamic convection conserves the kinetic energy 
and the magnetic convection conserves the magnetic energy, 
while the magnetic tension and the magnetic stretching 
convey the energy between the kinetic and magnetic energies. 

We apply the EDQNM approximation for two-dimensional case in the 
absence of cross helicity. 
Then the evolution equations of kinetic energy spectrum $E_V(k)$ 
and magnetic energy spectrum $E_M(k)$ 
are calculated to be 
\begin{eqnarray}
& &\left\{{\partial \over {\partial t}} + 2 \nu k^2\right\} E_V(k)= \nonumber \\
& &\qquad{2 \over \pi} \int_\Delta \theta_{kpq}
 \left\{ T_{vv}(k,p,q) +T_{BB}(k,p,q) \right\} dpdq,
\label{eq:EVk}
\end{eqnarray}
\begin{eqnarray}
& &\left\{{\partial \over {\partial t}} + 2 \eta k^2\right\} E_M(k)= \nonumber \\
& &\qquad{2 \over \pi} \int_\Delta \theta_{kpq}
 \left\{ T_{vB}(k,p,q) +T_{Bv}(k,p,q) \right\} dpdq,
\label{eq:EMk}
\end{eqnarray}
where
\begin{eqnarray}
T_{vv}(k,p,q) &=& 
{\sin \alpha \over k^2pq}(k^2-p^2)(p^2-q^2) \nonumber \\
& &\times \left\{ kE_V(p)E_V(q) - pE_V(q)E_V(k) \right\}, 
\nonumber
\end{eqnarray}
\begin{eqnarray}
T_{BB}(k,p,q) &=&
{p\sin \alpha \over k^2q}(q^2-p^2) \nonumber \\
& &\times\left\{ pE_M(k)E_M(q) - kE_M(p)E_M(q) \right\},
\nonumber
\end{eqnarray}
\begin{eqnarray}
& &T_{vB}(k,p,q) =
 {1 \over q \sin \alpha}
\left\{ {k^3 \over p}(y^2+xyz)E_V(p)E_M(q) \right. \nonumber \\
& &\quad+[ q^2(xyz+x^2y^2) -k^2(xyz+y^2z^2) ]E_M(k)E_M(q) \nonumber \\
& &\qquad \qquad\qquad\qquad
\left. -p^2(z^2+xyz)E_V(q)E_M(k) { { } \over { } }\right\},\nonumber
\end{eqnarray}
\begin{eqnarray}
T_{Bv}(k,p,q) &=&
 {1 \over q \sin \alpha}
\left\{  {k^3 \over p}(z^2+xyz)E_V(p)E_M(q) \right. \nonumber \\
& &+p^2(xyz+2x^2z^2-z^2)E_M(k)E_M(q) \nonumber \\
& &\left.-p^2(y^2+xyz)E_V(q)E_M(k) { { } \over { } }\right\}, \nonumber
\end{eqnarray}
are the nonlinear interactions corresponding to the terms
of hydrodynamic convection, magnetic tension,
magnetic convection and magnetic stretching 
in eqs.~(\ref{eq:NS}) and (\ref{eq:Maxwell}),
and $\theta_{kpq}$ is the triad-relaxation time, 
which depends on time and involves the eddy dumping rate 
for EDQNM approximation\cite{Pouquet}
or is equated to be constant for Markovian random coupling model\cite{Frisch}
(see refs.~7-10  
for the details).
The integral in the $pq$ plane is taken over the domain $\Delta$
that $k$, $p$ and $q$ form a triangle, and
$x$, $y$ and $z$ denote the cosines of the angles opposite to
the sides $k$, $p$ and $q$, respectively,
and $\alpha$ is the interior angle opposite to the side $k$.
We should remark that the nonlinear interaction reduced from
the total pressure term vanishes in the Fourier space 
and that $T_{vv}+T_{BB}$ and $T_{vB}+T_{Bv}$ are studied 
by Pouquet.\cite{Pouquet} 

We consider the effect of small-scale turbulence
on the large-scale field in terms of the eddy viscosities.
The eddy viscosities are obtained
by calculating the contribution of nonlinear interaction
with large wavenumbers $p$ and $q > k_m >>k$
in the evolution equations (\ref{eq:EVk}) and (\ref{eq:EMk}) 
of energy spectra $E_V(k)$ and $E_M(k)$. 
Then the rates of the change of the energy spectra
by the small-scale nonlinear interaction
can be written as
$$ {{\partial E_V(k)} \over {\partial t}} =
 - (\nu_{vv} + \nu_{BB} +2\nu ) k^2 E_V(k), $$
$$ {{\partial E_M(k)} \over {\partial t}} =
 - (\nu_{vB} + \nu_{Bv} +2\eta ) k^2 E_M(k), $$
where
\begin{eqnarray}
\nu_{vv} &=& {1 \over 4}
 \int^{\infty}_{k_m} \theta_{kpp} {\partial \over \partial p}
(pE_V(p))dp, \nonumber \\
\nu_{BB} &=& \int^{\infty}_{k_m} \theta_{kpp}
\left\{ E_M(p) -{1 \over 4}{\partial (pE_M(p)) \over \partial p} \right\} dp,
\nonumber \\
\nu_{vB}&=&\int^{\infty}_{k_m} \theta_{kpp}
\left\{ 2 E_V(p) -E_M(p) { { } \over { } } \right. \nonumber \\
& &
\qquad \left. -{1 \over 4}{\partial \over \partial p} \{p(E_V(p)+E_M(p))\} 
\right\} dp,
\nonumber \\
\nu_{Bv} &=& \int^{\infty}_{k_m} \theta_{kpp} \left\{ 
-E_M(p) { { } \over { } } \right. \nonumber \\
& &
\qquad \left. +{1 \over 4}{\partial \over \partial p} \{p(E_V(p)+E_M(p))\} 
\right\} dp,
\nonumber
\end{eqnarray}
are the eddy viscosities due to the nonlinear interactions
$T_{vv}$, $T_{BB}$, $T_{vB}$ and $T_{Bv}$, respectively.
The eddy viscosity of hydrodynamic convection term, $\nu_{vv}$,
is the same as the one for non-MHD 2-D turbulence
and it is negative since $E_V(p)$ decreases faster than
$1/p$ for large $p$.\cite{Ishizawa}
This implies the inverse cascade by this term.\cite{Pouquet}
The viscosity $\nu_{BB}$, which is due to the magnetic tension term, 
represents the effect of small-scale magnetic field
on the large-scale velocity field,
and is positive if $E_M(p)$ decreases faster than
$1/p$ for large $p$.\cite{Pouquet}

We have obtained the two eddy viscosities, $\nu_{vB}$ and $\nu_{Bv}$, 
which are due to 
the magnetic convection and the magnetic stretching term, respectively.
Note that these viscosities are different from those considered in 
Pouquet,\cite{Pouquet} who decomposed $\nu_{vB}+\nu_{Bv}$ by the 
type of acting energy. 
The eddy viscosity $\nu_{vB}$ 
is positive if $2E_V(p) -E_M(p)>0$ for $p > k_m$
and total energy, $E_V(p)+E_M(p)$, decreases faster than $1/p$.
These conditions are satisfied in our numerical results 
for $k_m$ larger than
the wave number in the middle of inertial range\cite{Ishizawa}; 
thus the magnetic convection yields the normal cascade.
The eddy viscosity $\nu_{Bv}$ is negative, 
$$ \nu_{Bv} < 0, $$
since the total energy
decreases faster than $1/p$.
Therefore
the magnetic stretching term transports the energy
from the small scale to the large scale
to form the large coherent magnetic structures.
It appears to be contradictory 
that although the total eddy viscosity $\nu_{vB}+\nu_{Bv}$
for magnetic energy spectrum, 
already obtained by Pouquet,\cite{Pouquet} 
is negative for our numerical results 
which show $E_V(p) -E_M(p) <0$ for $p > k_m$,\cite{Ishizawa} 
the numerical results exhibit the energy transfer of
the magnetic energy to the small scale.\cite{Ishizawa} 
This point should be studied further. 
We remark here that the inequality $E_V(p) -E_M(p) <0$ for $p > k_m$ seems 
not to be generally true.\cite{Ting} 

Using the EDQNM approximation, 
we have shown that
the magnetic stretching term
is responsible for
turning 
the small-scale turbulent structures
into
the large coherent magnetic structures
in the 2-D MHD turbulence.
This mechanism, first deduced from numerical results for a single 
set of parameters, has now a theoretical background and 
it is expected to be an important mechanism in the 
2-D MHD turbulence generally. 
In the process of fast coalescence of small magnetic eddies,
fine structures of magnetic field may be removed
by this term through the eddy viscosity.\cite{Ishizawa}

One of the authors (A.~I.) would like to thank Dr.~M.~Azumi
for his encouragement.

\end{document}